\documentclass[twocolumn]{el-author}
\date{14th August 2014}

\begin{document}

\title{Fast Identification of Wiener-Hammerstein Systems using Discrete Optimization}

\author{M. Schoukens, G. Vandersteen, Y. Rolain, and F. Ferranti}

\abstract{This letter proposes a fast identification algorithm for Wiener-Hammerstein systems. The computational cost of separating the front and the back linear time invariant block dynamics is significantly improved by using discrete optimization. The discrete optimization is implemented as a genetic algorithm. Numerical results confirm the efficiency and accuracy of the proposed approach.}

\maketitle

\section{Introduction}
\begin{figure}
		\centering
			\includegraphics[width=0.80\columnwidth]{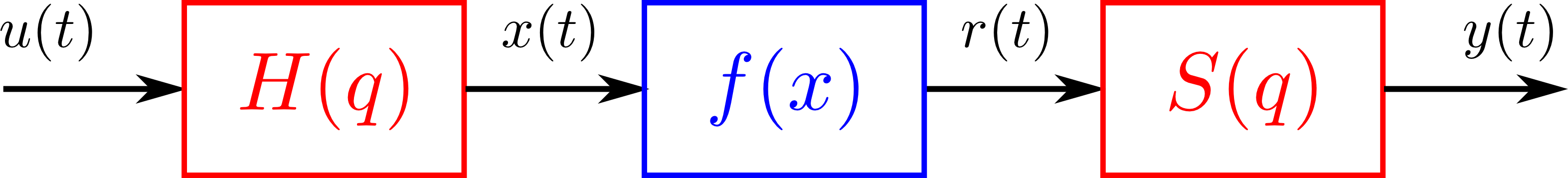}
		\caption{A Wiener-Hammerstein system: a static nonlinear block $f(x)$ sandwiched in between two LTI blocks $H(q)$ and $S(q)$.}
		\label{fig:WienerHammerstein}
	\end{figure}

	Block-oriented structures are useful to model a nonlinear system. Applications range from RF amplifiers over chemical processes to physiological systems \cite{Giri2010}. A block-oriented model consists of two types of blocks: Linear Time Invariant (LTI) and static nonlinear blocks. The most simple block-oriented model structures are the Wiener (a LTI block followed by a static nonlinear block) \cite{Giri2010},\cite{Billings1977} and the Hammerstein (static nonlinear block followed by a LTI block) \cite{Giri2010},\cite{Crama2004} model structure. A straightforward extension of the Wiener and the Hammerstein structure is the Wiener-Hammerstein structure. A Wiener-Hammerstein model structure (see Figure \ref{fig:WienerHammerstein}) is given by a static nonlinearity that is sandwiched in between two LTI blocks \cite{Giri2010},\cite{Sjoberg2012},\cite{Vanbeylen2014}. In this letter we assume that the output $y(t)$ of a Wiener-Hammerstein system is given by:
	\begin{align}
		y(t) = S(q)\left[ f\left( H(q)\left[ u(t)\right] \right) \right], \label{eq:ModelOutput1}
	\end{align}	
	where $q^{-1}$ denotes the backwards shift operator and $u(t)$ is the noise free input signal. $H(q)$ and $S(q)$ are rational forms in the backward shift operator $q^{-1}$, and $f(x)$ is assumed here to consist of a linear combination of basis functions for the nonlinearity (e.g. polynomials):
	\begin{align} 
		H(q) &= \frac{D(q)}{C(q)} 
				 = \frac{d_{0} + d_{1}q^{-1} + \ldots + d_{n_d}q^{-n_d}}{c_{0} + c_{1}q^{-1} + \ldots + c_{n_c}q^{-n_c}},	 \label{eq:LTIh}	\\
		S(q) &= \frac{B(q)}{A(q)}
				 = \frac{b_{0} + b_{1}q^{-1} + \ldots + b_{n_b}q^{-n_b}}{a_{0} + a_{1}q^{-1} + \ldots + a_{n_a}q^{-n_a}},  \label{eq:LTIs}		\\
		r(t) &= f(x(t)) = \sum_{j=1}^{n_w}w_{j}f_{j}(x(t)). \label{eq:StaticNL}
	\end{align}
	
	Obtaining an estimate for the two LTI-blocks is the main difficulty during the identification of a Wiener-Hammerstein system. A brute-force approach is proposed in \cite{Sjoberg2012} to split the dynamics that are present at the front and at the back of the Wiener-Hammerstein structure. This method is quite straightforward and obtains good results. Its main drawback is the high computational time that is needed for the brute-force scan. 
	
	A fractional power approach is proposed in \cite{Vanbeylen2014} to avoid the time-consuming scanning procedure. All the poles and zeros are present in both subsystems $H(q)$ and $S(q)$ with an exponent of respectively $\alpha$ and $1-\alpha$ ($0 \leq \alpha \leq 1$):
		\begin{align}
			H(q) &= \frac{D(q)}{C(q)} = \frac{\prod_{i=1}^{n_z}(z_i-q^{-1})^{\alpha_{z,i}}}{\prod_{i=1}^{n_p}(p_i-q^{-1})^{\alpha_{p,i}}},  						\\
			S(q) &= \frac{B(q)}{A(q)} = \frac{\prod_{i=1}^{n_z}(z_i-q^{-1})^{(1-\alpha_{z,i})}}{\prod_{i=1}^{n_p}(p_i-q^{-1})^{(1-\alpha_{p,i})}}, 			
		\end{align}	
	where $n_p$ and $n_z$ denote the number of poles and the number of zeros respectively. This replaces the brute-force scan by a continuous (constrained) optimization over the fractional powers $\alpha_{p,i}$ and $\alpha_{z,i}$ of the poles and zeros. This allows the poles and zeros to shift smoothly from $H(q)$ to $S(q)$. The main disadvantage of this optimization scheme is that it results in a fractional power representation of $H(q)$ and $S(q)$, which can be a local optimum. This fractional power representation can be recast into a power representation as in \eqref{eq:LTIh} and \eqref{eq:LTIs} in a second optimization step, but it makes the method more computationally involved.
	
	This letter proposes to use a discrete optimization approach to speed up the identification process, while keeping a high accuracy level and the simplicity of the brute-force approach. The outline of the letter is as follows. The first section summarizes the method that is presented in \cite{Sjoberg2012}. Next, the discrete optimization approach is explained. Finally, a simulation example shows the improvement in accuracy and efficiency that is obtained using the discrete optimization approach.
	
\section{Brute-force Identification by Pole-Zero Allocation Scan} \label{sec:BruteForce}
	The algorithm proposed in \cite{Sjoberg2012} starts with the estimation of the poles and zeros of the overall dynamics that are present in the nonlinear system. This is done using the Best Linear Approximation (BLA) \cite{Pintelon2012}. When input signals belonging to the Riemann equivalence class of asymptotically normally distributed excitation signals are used, the BLA of a Wiener-Hammerstein system is given by \cite{Pintelon2012}:
	\begin{align}
		G_{BLA} = \alpha H(q)S(q),
	\end{align}
	where $\alpha$ is a gain factor that depends on the power spectrum of the input signal and the Wiener-Hammerstein system under study. The poles and the zeros of the parametrized BLA are also the poles and zeros of the LTI blocks $H(q)$ and $S(q)$ of the Wiener-Hammerstein system.
	
	These poles and zeros of the BLA need to be assigned to either $H(q)$ or $S(q)$. All possible partitions of poles/zeros $\hat{H}_k(q)$ and $\hat{S}_k(q)$ are created such that:
	\begin{align}
		G_{BLA} = \alpha_k \hat{H}_k(q)\hat{S}_k(q),
	\end{align}
	where $k$ is the combination index and $\alpha_k$ is a gain factor, which is included in the estimation of the static nonlinearity.
	
	Next, a static nonlinearity is estimated for every combination $k$. This problem is linear in the parameters if the static nonlinearity $\hat{f}_k(x(t))$ is described by a linear combination of basis functions. It is solved using linear least squares regression, resulting in the estimate $\hat{f}_k$.
	
	Finally, all the estimated models are ranked based on their mean squares error $e_k$:
		\begin{align}
		 e_k &= \frac{1}{N} \sum_{t=1}^{N} \left(y(t) - \hat{y}_k(t)\right)^2, \\
		 \hat{y}_k(t) &= \hat{S}_k(q)\left[ \hat{f}_k\left( \hat{H}_k(q)\left[ u(t)\right] \right) \right],
		\end{align}
	where $N$ is the number of data points in the measured input-output record. The model with the lowest error is selected.
	
	The total number of combinations $k$ depends on the number of poles and the number of zeros, $n_p$ and $n_z$ respectively. It equals the total number of least squares regressions that needs to be performed. The total number of combinations is minimum $2^{\frac{n_{p}+n_{z}}{2}}$ and maximum $2^{n_{p}+n_{z}}$. The minimum number of combinations is obtained when all poles and zeros are part of a complex conjugate pair, the maximum number of combinations is obtained when all poles and zeros are real. This number increases very rapidly with the model order, which makes the brute-force scan approach in \cite{Sjoberg2012} computationally expensive. It is important to note that each complex conjugate pole or zero pair is assigned either to $S(q)$ or $H(q)$ in the combinations. The single elements of the pair are never assigned separately. This allows the LTI blocks to be transfer functions with real coefficients.
	
	A similar approach is also used for the identification of parallel Wiener-Hammerstein systems \cite{SchoukensM2013}. A parallel Wiener-Hammerstein system consists of several Wiener-Hammerstein subsystems that share the same input. The output of a parallel Wiener-Hammerstein system is given by the sum of the outputs of the Wiener-Hammerstein subsystems. The computational cost of the pole-zero partitioning step is even higher in this case since the brute-force scan is performed for all the parallel branches simultaneously.
	
\section{Pole-Zero Allocation using Discrete Optimization} \label{sec:DiscreteOptimization}
	The discrete optimization-based approach proposed in this letter speeds up the process of finding the best pole and zero allocation in a Wiener-Hammerstein model. The position of each pole and zero is represented by a binary value. The value $1$ indicates that the pole or zero belongs to the front LTI block $H(q)$, the value $0$ that it belongs to the back LTI block $S(q)$. This results in a binary parameter vector $\boldsymbol{\theta}$. The optimization of the binary parameter vector is performed using a genetic algorithm. Genetic algorithms belong to the larger class of evolutionary algorithms, which generate solutions to optimization problems using techniques inspired by natural evolution, such as mutation, selection and crossover \cite{Back1996},\cite{Ahn2006}. Genetic algorithms are used in several fields such as bioinformatics, computational science, engineering, economics, chemistry, mathematics and physics. The static nonlinearity is estimated each time the cost function is evaluated for a certain value of the parameter vector $\Theta$, as explained in the previous section.
	
	Instead of evaluating all possible pole/zero combinations, only a subset will be evaluated using the genetic algorithm. This results in a significant speedup of the identification algorithm.
	
%	Note that the proposed discrete optimization approach can easily be extended for parallel Wiener-Hammerstein systems \cite{SchoukensM2013}.

\section{Simulation Example} \label{sec:Simulation}
	A Monte Carlo simulation is performed to show the good results in accuracy and efficiency that are obtained using the discrete optimization approach. In each simulation a Wiener-Hammerstein system is created with a polynomial static nonlinearity of degree 3:
	\begin{align}
		r = 3x + w_2 x^2 + w_3 x^3.
	\end{align}
	The coefficient of the linear term is equal to 3, the 2nd and 3rd degree coefficients $w_2$ and $w_3$ are uniformly distributed in the range $[-0.25,0.25]$. The simulation is performed for different LTI orders. The front LTI block and the back LTI block are Chebychev filters of the same order ($n_a = n_b = n_c = n_d = [5,6,7,8]$). The front LTI block is a Chebychev type 1 filter with an in band ripple of 3 dB and a cut-off frequency that is uniformly distributed in the range $[0.025 f_s,0.125 f_s]$. The back LTI block is a Chebychev type 2 filter with a stop band ripple at 50 dB and a cut-off frequency that is uniformly distributed in the range $[0.025 f_s,0.125 f_s]$. The cut-off frequencies of the LTI blocks are independent of each other.
	
	The brute-force algorithm and the discrete optimization approach both use the same cost function. The brute-force algorithm evaluates this cost function for every possible parameter vector $\boldsymbol{\theta}$. The discrete optimization approach evaluates this cost function for a much smaller set of parameter vectors. The genetic algorithm that is present in the Matlab Optimization toolbox ('ga' function name) is used. The optimization settings used for the simulations are shown in Table \ref{tab:GaOptions}. An important setting of the genetic algorithm is the population size. If the population size is too small, the algorithm might get stuck in a local optimum, a too large population size slows down the algorithm since it needs to perform many function evaluations.
	
	The true zeros and poles of the LTI blocks are used, instead of the zeros and poles of the estimated BLA, to simplify the Monte Carlo simulation. The cost function is evaluated with 1 period of a periodic Gaussian noise signal with a standard deviation equal to one. One period contains 4096 points. 100 simulations are performed for each model order. The low degree of the nonlinearity in the system, and the limited number of data points that is used during the cost function evaluation ensure that the Monte Carlo simulation can be performed in a reasonable amount of time.
	
\begin{table} 
	\caption{Settings of the Matlab Optimization toolbox function 'ga'.}
	\centering
		\begin{tabular}{c|c}
			Setting Name	 & setting \\ \hline
			PopulationType & bitstring \\
			CreationFcn		 & gacreationuniform \\
			CrossoverFcn	 & crossoverscattered \\
			MutationFcn		 & mutationuniform \\
			Generations		 & 50 \\
			PopulationSize & see Table \ref{tab:SimResults} \\
			StallGenLimit	 & 5 \\
			TolFun		     & 1e-20 \\
		\end{tabular}
		\label{tab:GaOptions}
\end{table}

\begin{table} 
	\caption{Monte Carlo simulation results: average time and success rates.}
	\centering
		\begin{tabular}{c|c|c|c|c}
			$n_a = n_b = n_c = n_d$						& 5 			& 6					& 7				& 8 					\\ \hline
%			Avg. \# Combinations to Scan		&	4096		&	6103.04		&	65536		&	110100.48		\\
			Population Size 									& 200			& 400				& 600			& 800 				\\
			brute-force Scan									& 8.73s 	& 12.9s			& 149s		& 247s				\\
			Discrete optimization (GA)				& 4.12s 	& 7.30s			& 14.7s		& 18.9s				\\
			Success Rate											& 98\%		& 97\%			& 95\%		& 95\%				\\
		\end{tabular}
		\label{tab:SimResults}
\end{table}
	
	The discrete optimization approach clearly speeds up the identification process, as is shown in Table \ref{tab:SimResults}. A speed-up factor equal to 10 is obtained for the higher model orders, while a speed-up factor equal to 2 (a bit less than 2 for the model order 12 case) is obtained for the lower model orders. The success rate shows the percentage of the cases where the discrete optimization and the brute-force scan ended up in the same minimum value of the cost function. A success rate of 95\% to 100\% is achieved in all cases. An even higher speed-up can be achieved, but this comes at the cost of a lower succes rate. 
	
\section{Conclusion} \label{sec:Conclusion}
	
	We have presented a fast identification algorithm for Wiener-Hammerstein systems. The problem is reformulated such that a binary parameter vector can be used to represent the combinations of poles and zeros in the front and the back LTI blocks. A genetic algorithm is used to perform the discrete optimization. The simulation results show that the proposed method is able to identify a Wiener-Hammerstein system ten times faster than the brute-force scanning approach for the high order models. The proposed method is still almost twice as fast as the brute-force method for lower order models. The proposed method is more efficient than the brute-force scan approach, while it achieves a very similar accuracy level.
	The extension of the proposed algorithm towards parallel Wiener-Hammerstein systems will be investigated as future work.

\vskip3pt
\ack{This work was supported in part by the VUB (SRP-19), the Fund for Scientific Research (FWO-Vlaanderen), the Methusalem grant of the Flemish Government (METH-1), by the Belgian Government through the Inter university Poles of Attraction IAP VII/19 DYSCO program, and the ERC advanced grant SNLSID, under contract 320378. F. Ferranti is currently a Post-Doctoral Research Fellow of FWO-Vlaanderen. M. Schoukens is currently an FWO Aspirant, supported by FWO-Vlaanderen.}

\vskip5pt

\noindent M. Schoukens, G. Vandersteen, and Y. Rolain (\textit{Department Elec, Vrije Universiteit Brussel (VUB), Pleinlaan 2, B-1050 Brussel, Belgium})

\noindent F. Ferranti (\textit{Department of Information Technology, Ghent University-iMinds, Gaston Crommenlaan 8/201, B-9050 Gent, Belgium})

\vskip3pt

\noindent E-mail: maarten.schoukens@vub.ac.be


\begin{thebibliography}{}

\bibitem{Giri2010}
Giri, F. and Bai, E.W. (editors): `Block-oriented Nonlinear System Identification', \textit{Volume 404 of Lecture Notes in Control and Information Sciences}, 2010, Berlin, Heidelberg.

\bibitem{Billings1977}
Billings, S.A. and Fakhouri, S.Y.: `Identification of nonlinear systems using the Wiener model', \textit{Electronics Letters}, 1977, \textbf{13}(17), pp. 502-504.

\bibitem{Crama2004}
Crama, P., Schoukens, J. and Pintelon, R.: 'Generation of enhanced initial estimates for Hammerstein systems', \textit{Automatica}, 2004, \textbf{40}(7), pp. 1269-1273.

\bibitem{Sjoberg2012}
Sj\"oberg, J. and Schoukens, J.: `Initializing Wiener-Hammerstein models based on partitioning of the best linear approximation', \textit{Automatica}, 2012, \textbf{48}(2), pp. 353-359.

\bibitem{Vanbeylen2014}
Vanbeylen, L.: `A fractional approach to identify Wiener-Hammerstein systems', \textit{Automatica}, 2014, \textbf{50}(3), pp. 903-909.

\bibitem{Pintelon2012}
Pintelon, R. and Schoukens, J.: 'System Identification: A Frequency Domain Approach', \textit{Wiley-IEEE Press}, 2012, Hoboken, New Jersey.

\bibitem{SchoukensM2013}
Schoukens, M., Vandersteen, G. and Rolain, Y.: 'An identification algorithm for parallel Wiener-Hammerstein systems', \textit{52nd IEEE Conference on Decision and Control (CDC)}, 2013, 4907-4912.

\bibitem{Back1996}
B\"ack, T.: 'Evolutionary algorithms in theory and practice evolution strategies, evolutionary programming, genetic algorithms', \textit{Oxford University Press}, 1996, New York.

\bibitem{Ahn2006}
Ahn, C.W.: 'Advances in evolutionary algorithms theory, design and practice', \textit{Springer}, 2006, Berlin, New York.

\end{thebibliography}
\end{document}